\documentclass{revtex4}
\usepackage{amssymb}
\usepackage{amsfonts}
\usepackage{amsmath}

\input{tcilatex}

\begin{document}

\title{(1+1)-Dirac bound states in one-dimension; position-dependent Fermi
velocity and mass}
\author{Omar Mustafa}
\email{omar.mustafa@emu.edu.tr}
\affiliation{Department of Physics, Eastern Mediterranean University, G. Magusa, north
Cyprus, Mersin 10 - Turkey,\\
Tel.: +90 392 6301378; fax: +90 3692 365 1604.}

\begin{abstract}
\textbf{Abstract: }We extend Panella and Roy's \cite{13} work for massless
Dirac particles with position-dependent (PD) velocity. We consider Dirac
particles where the mass and velocity are both position-dependent. Bound
states in the continuum (BIC)-like and discrete bound state solutions are
reported. It is observed that BIC-like solutions are not only feasible for
the ultra-relativistic (massless) Dirac particles but also for Dirac
particles with PD-mass and PD-velocity that satisfy the condition $m\left(
x\right) v_{F}^{2}\left( x\right) =A$, where $A\geq 0$ is constant. A Dirac P%
\"{o}schl-Teller and a harmonic oscillator models are also reported.


\end{abstract}

\maketitle

\section{Introduction}

In heterostructure physics, it was believed that electrons are effectively
described by the position-dependent mass (PDM) Schr\"{o}dinger Hamiltonian
(i.e., von Roos Hamiltonian, e.g., \cite{1,2,3} ). Using the Pauli spin
matrices in the Schr\"{o}dinger Hamiltonian, the Dirac Hamiltonian was
ignored. However, this perspective has drastically changed since the
discovery of graphene \cite{4,5}. Many studies on the applicability of Dirac
Hamiltonian in condense matter were carried out (cf., e.g., \cite%
{4,5,6,7,8,9,10,11,12,13} and related references cited therein) . It is
found that the effective low-energy model for the quasi-particles is
ultrarelativistic (i.e., massless) and is described by the Hamiltonian%
\begin{equation}
H=v_{F}\,\mathbf{\sigma \cdot p.}
\end{equation}%
Which is in fact the Dirac Hamiltonian for massless particles with an
effective Fermi velocity $v_{F}$ (where $v_{F}=c/300$, $c$ is the speed of
light, $\mathbf{\sigma }$ is a vector using Pauli matrices and $\mathbf{p}=-i%
\vec{\nabla}$, with $\hbar =1$).

However, the information on the material properties may be encoded in the
Fermi velocity of the Dirac particles \cite{6,7}. In this case, the Dirac
Hamiltonian (1) takes the form%
\begin{equation}
H=\sqrt{v_{F}\left( x\right) }\sigma _{x}p_{x}\sqrt{v_{F}\left( x\right) }.
\end{equation}%
Hereby, one should notice that the replacement of the constant velocity, $%
v_{F}$, by the position-dependent one, $v_{F}\left( x\right) $, would render
Hamiltonian (1) non-Hermitian. Whereas the form of Hamiltonian (2) preserves
Hermiticity and recovers the constant $v_{F}$ setting. Panella and Roy \cite%
{13}, for example, have used Hamiltonian (2) to study bound states in the
continuum (BIC) (cf., e.g., \cite{13} and related references therein) and
discrete energy states for massless Dirac particle. Throughout this paper,
we shall refer to their study as Panella-Roy's model (namely, their model
with $m\left( x\right) =0$ and $v_{F}\left( x\right) =v_{0}\cosh ^{2}\alpha
x $). They have found that with proper PD-Fermi velocity profile it is
possible to create BIC-like and discrete bound-state solutions.

In this paper, motivated by theoretical curiosity and/or possible practical
applicability, we propose that the information on the material properties is
not only encoded in the Fermi velocity but also encoded in the mass of the
Dirac particles. We therefore extend Panella and Roy's \cite{13} work and
consider the Dirac-Hamiltonian where the Fermi velocity and the mass are
both position dependent. That is, we shall work with the Hamiltonian%
\begin{equation}
H=\sqrt{v_{F}\left( x\right) }\sigma _{x}p_{x}\sqrt{v_{F}\left( x\right) }%
+\beta m\left( x\right) v_{F}\left( x\right) ^{2},
\end{equation}%
where $\sigma _{x}$ and $\beta $ are the usual Pauli matrices \cite{6,7}.
Moreover, it is obvious that the second term in (3) is analogous PDM Dirac
particle in a Lorentz scalar potential (cf., e.g., \cite{14} and related
references therein). The addition of such term leaves the corresponding
Dirac Hamiltonian invariant under Lorentz transformation. The organization
of this paper is in order.

We discuss Hamiltonian (3) and give our methodical proposal in section II.
We provide illustrative examples, including ultra-relativistic Dirac
quasi-particles (i.e., particles with $m\left( x\right) =0$), in section
III. In the same section, we show that similar scenarios (as those in the
Panella-Roy's model \cite{13} for BIC-like and for discrete bound-states
solutions) are observed for a wider class of $m\left( x\right) $ and $%
v_{F}\left( x\right) $ (i.e., for $m\left( x\right) v_{F}^{2}\left( x\right)
=A$, where $A\geq 0$ is constant). For such mass and Fermi velocity
settings, a shift in the energy levels is obtained. In section IV, we show
that Dirac particles may be trapped in an effective P\"{o}schl-Teller
potential \cite{15} produced by both their PD-mass and PD-Fermi velocity.
Moreover, we show, in section V, that the (1+1)-Dirac oscillator may just be
a consequence of a linear PD-Fermi velocity and a singular PD-mass (i.e., $%
v_{F}\left( x\right) =v_{0}x$ and $m\left( x\right) =A/x$). Our concluding
remarks are given in section VI.

\section{(1+1)-Dirac particles with position-dependent velocity and mass}

With the usual textbook Pauli matrices and Dirac spinors, the (1+1)-Dirac
equation $H\psi \left( x\right) =E\psi \left( x\right) $, for $H$ in (3),
would decouple into%
\begin{eqnarray}
-iF\left( x\right) \partial _{x}\left[ F\left( x\right) \psi _{2}\left(
x\right) \right] &=&\zeta _{1}\left( x\right) \psi _{1}\left( x\right) \text{
; \ }\zeta _{1}\left( x\right) =E-m\left( x\right) F\left( x\right) ^{4}, \\
-iF\left( x\right) \partial _{x}\left[ F\left( x\right) \psi _{1}\left(
x\right) \right] &=&\zeta _{2}\left( x\right) \psi _{2}\left( x\right) \text{
; \ }\zeta _{2}\left( x\right) =E+m\left( x\right) F\left( x\right) ^{4},
\end{eqnarray}%
where $F\left( x\right) =\sqrt{v_{F}\left( x\right) }$ and%
\begin{equation}
\psi \left( x\right) =N\left( 
\begin{tabular}{l}
$\psi _{1}\left( x\right) $ \\ 
$\psi _{2}\left( x\right) $%
\end{tabular}%
\right) ,
\end{equation}%
(with $N$ as the normalization constant) are used. Now let us multiply (4)
and (5), from the left, by $F\left( x\right) $ and use the substitutions $%
\tilde{\psi}_{1,2}\left( x\right) =F\left( x\right) \psi _{1,2}\left(
x\right) $ to imply%
\begin{equation}
\tilde{\psi}_{2}\left( x\right) =-i\left( \frac{F\left( x\right) ^{2}}{\text{%
\ }\zeta _{2}\left( x\right) }\right) \partial _{x}\tilde{\psi}_{1}\left(
x\right) .
\end{equation}%
Which when substituted in (4) yields%
\begin{equation}
-\partial _{x}^{2}\tilde{\psi}_{1}\left( x\right) +\left[ \frac{\zeta
_{2}^{\prime }\left( x\right) }{\zeta _{2}\left( x\right) }-2\left( \frac{%
F^{\prime }\left( x\right) }{F\left( x\right) }\right) \right] \partial _{x}%
\tilde{\psi}_{1}\left( x\right) =\left( \frac{\zeta _{1}\left( x\right)
\zeta _{2}\left( x\right) }{F\left( x\right) ^{4}}\right) \tilde{\psi}%
_{1}\left( x\right) ,
\end{equation}%
where primes denote derivatives with respect to $x$. To get rid of the first
order derivative and bring (8) into the one-dimensional form of Schr\"{o}%
dinger equation we use%
\begin{equation}
\tilde{\psi}_{1}\left( x\right) =\zeta _{2}\left( x\right) ^{\nu }\phi
_{1}\left( q\right) \text{ ; \ }q\equiv q\left( x\right) .
\end{equation}%
This would suggest that%
\begin{equation}
q^{\prime }\left( x\right) =\zeta _{2}\left( x\right) ^{1-2\nu }F\left(
x\right) ^{-2}.
\end{equation}%
However, one also needs to avoid position-dependent energies and choose $\nu
=1/2$ to imply%
\begin{equation}
-\partial _{q}^{2}\phi _{1}\left( q\right) +v_{F}\left( x\right) ^{2}\left[ 
\frac{3}{4}\left( \frac{\zeta _{2}^{\prime }\left( x\right) }{\zeta
_{2}\left( x\right) }\right) ^{2}-\frac{1}{2}\frac{\zeta _{2}^{\prime \prime
}\left( x\right) }{\zeta _{2}\left( x\right) }-\frac{1}{2}\left( \frac{%
v_{F}^{\prime }\left( x\right) }{v_{F}\left( x\right) }\right) \frac{\zeta
_{2}^{\prime }\left( x\right) }{\zeta _{2}\left( x\right) }\right] \phi
_{1}\left( q\right) =\left[ E^{2}-m\left( x\right) ^{2}v_{F}\left( x\right)
^{4}\right] \phi _{1}\left( q\right) ,
\end{equation}%
with%
\begin{equation}
q\left( x\right) =\int\nolimits^{x}\frac{1}{v_{F}\left( y\right) }dy,
\end{equation}%
where $q\left( x\right) $ represents a point canonical transformation. It is
obvious that for massless particles equation (11) collapses into its most
simplistic form%
\begin{equation*}
-\partial _{q}^{2}\phi _{1}\left( q\right) =E^{2}\phi _{1}\left( q\right) ,
\end{equation*}%
that looks very much like the one-dimensional Schr\"{o}dinger equation for
free particles. However, the form of $v_{F}\left( x\right) $ would determine
the domain of $q\left( x\right) $ in (12) and, therefore, has its say in the
process. This is to be clarified in the illustrative examples below.

Nevertheless, one may use the non-relativistic limit $m\left( x\right)
v_{F}\left( x\right) ^{2}>>E_{binding}\equiv E_{bind}$, where $%
E_{bind}=E-m\left( x\right) v_{F}\left( x\right) ^{2}$ (analogous to the
textbook non-relativistic limit for Dirac particles with rest mass energy $%
m_{\circ }c^{2}>>E_{bind}$ \cite{16}). One would, in this way, recover the
constant non-zero mass and constant velocity settings as well as accommodate
position-dependent mass $m\left( x\right) $ at $v_{F}\left( x\right) =c$.
This non-relativistic limit would, in turn, yield%
\begin{equation}
\frac{1}{\zeta _{2}\left( x\right) }=\frac{1}{E+m\left( x\right) v_{F}\left(
x\right) ^{2}}=\frac{1}{E_{bind}+2m\left( x\right) v_{F}\left( x\right) ^{2}}%
\approx \frac{1}{2m\left( x\right) v_{F}\left( x\right) ^{2}}.
\end{equation}%
Consequently, one may recast (11) as%
\begin{equation}
-\partial _{q}^{2}\phi _{1}\left( q\right) +V_{eff}\left( q\right) \phi
_{1}\left( q\right) =E^{2}\phi _{1}\left( q\right)
\end{equation}%
where%
\begin{equation}
V_{eff}\left( q\right) =\frac{3}{16}\frac{\left[ \left( m\left( x\right)
v_{F}\left( x\right) ^{2}\right) ^{\prime }\right] ^{2}}{m\left( x\right)
^{2}v_{F}\left( x\right) ^{2}}-\frac{1}{4}\left[ \frac{\left( m\left(
x\right) v_{F}\left( x\right) ^{2}\right) ^{\prime \prime }}{m\left(
x\right) }+\left( \frac{v_{F}^{\prime }\left( x\right) }{v_{F}\left(
x\right) }\right) \frac{\left( m\left( x\right) v_{F}\left( x\right)
^{2}\right) ^{\prime }}{m\left( x\right) }\right] +m\left( x\right)
^{2}v_{F}\left( x\right) ^{4}.
\end{equation}%
Obviously, this approximation may only be used for non-zero constant mass
and not for massless Dirac particles. To illustrate our methodical proposal
above we discuss the following illustrative examples.

\section{BIC-like and discrete bound-states solutions: parallel and
complementary to Panella-Roy's model}

One considers the class of \ PD-mass and PD-Fermi velocity satisfying the
condition $m\left( x\right) v_{F}\left( x\right) ^{2}=A$, where $A\geq 0$ is
a constant. Under such assumptions, equation (11) would read%
\begin{equation}
-\partial _{q}^{2}\phi _{1}\left( q\right) =\tilde{\lambda}^{2}\phi
_{1}\left( q\right) ,
\end{equation}%
where $\tilde{\lambda}^{2}=E^{2}-A^{2}$. Yet, one may notice that $m\left(
x\right) =0$ is just a special case of the current more general proposal
than that used in Panella-Roy's\ model \cite{13}. Although this equation
looks like Schr\"{o}dinger equation for free particle, the domain of $%
q\left( x\right) $ in (12) would determine the boundary conditions on the
related state functions. This is to be clarified in the following two
examples. The first of which is discussed here as a complementary model that
reports on the consequences of using $m\left( x\right) =m_{\circ }/\cosh
^{4}\alpha x$, \ and \ $v_{F}\left( x\right) =v_{0}\cosh ^{2}\alpha x$
settings on Panella-Roy's model. The second example considers $v_{F}\left(
x\right) =v_{0}\left( 1+\alpha ^{2}x^{2}\right) $, $m\left( x\right)
=m_{\circ }/(1+\alpha ^{2}x^{2})^{2}$ and shares similar scenario on the
BIC-like and the discrete bound-states solutions as that reported in
Panella-Roy's model \cite{13}.

\subsection{Consequences of $m\left( x\right) v_{F}\left( x\right) ^{2}=A$
on Panella-Roy's model: complementary}

Let the PD-mass and the PD-Fermi velocity take the following forms%
\begin{equation}
m\left( x\right) =\frac{m_{\circ }}{\cosh ^{4}\alpha x}\text{, \ and \ }%
v_{F}\left( x\right) =v_{0}\cosh ^{2}\alpha x
\end{equation}%
respectively, with the constants $m_{\circ },v_{0}\geq 0$ and $A=m_{\circ
}v_{0}^{2}$. This would, in turn, imply%
\begin{equation}
q\left( x\right) =\int\nolimits^{x}\frac{1}{v_{F}\left( y\right) }dy=\frac{1%
}{\alpha v_{0}}\tanh \alpha x,
\end{equation}%
and suggest that $q\left( x\right) \in \left( -1/\alpha v_{0},1/\alpha
v_{0}\right) $. Therefore, our particle under consideration is not free but
rather quasi-free (i.e., trapped in a force field produced by its own
PD-mass and PD-Fermi velocity in (17)) and is confined to move between $%
-1/\alpha v_{0}$ and $+1/\alpha v_{0}$.

Whilst the solution of (16) is straightforward and takes the form%
\begin{equation}
\phi _{1}\left( q\right) =\sin \tilde{\lambda}q,
\end{equation}%
it is rather unphysical (i.e., it does not satisfy the boundary conditions
imposed by the range of $q\left( x\right) $). Nevertheless, one may use this
unphysical solution to obtain the related un-normalized wave function
components through (9) and (7) as%
\begin{equation}
\psi _{1}\left( x\right) =\frac{\tilde{\psi}_{1}\left( x\right) }{\sqrt{%
v_{F}\left( x\right) }}=\sqrt{\frac{\zeta _{2}}{v_{0}}}\text{sech}\left(
\alpha x\right) \sin \left[ \frac{\tilde{\lambda}}{\alpha v_{0}}\tanh \alpha
x\right] \text{ ; }\zeta _{2}=E+m_{\circ }v_{0}^{2},
\end{equation}%
and%
\begin{equation}
\psi _{2}\left( x\right) =\frac{\tilde{\psi}_{2}\left( x\right) }{\sqrt{%
v_{F}\left( x\right) }}=-i\sqrt{\frac{\zeta _{1}}{v_{0}}}\text{sech}\left(
\alpha x\right) \cos \left[ \frac{\tilde{\lambda}}{\alpha v_{0}}\tanh \alpha
x\right] \text{ ; }\zeta _{1}=E-m_{\circ }v_{0}^{2}.
\end{equation}%
Under such settings, the probability density $\rho \left( x\right) $ is
given by%
\begin{equation}
\rho \left( x\right) =\left\vert \psi _{1}\left( x\right) \right\vert
^{2}+\left\vert \psi _{2}\left( x\right) \right\vert ^{2}=N^{2}\frac{\zeta
_{1}}{v_{0}}\text{sech}^{2}\left( \alpha x\right) +2m_{\circ }v_{0}^{2}\text{%
sech}^{2}\left( \alpha x\right) \cos ^{2}\left[ \frac{\tilde{\lambda}}{%
\alpha v_{0}}\tanh \alpha x\right] ,
\end{equation}%
and the normalization constant $N$ is obtained through%
\begin{equation}
\int\limits_{-\infty }^{\infty }\rho \left( x\right) dx=2N^{2}\left( \frac{%
\zeta _{1}}{\alpha v_{0}}+\frac{2m_{\circ }v_{0}^{3}}{\tilde{\lambda}}%
\right) =1\Longrightarrow N=\sqrt{\frac{\alpha v_{0}\tilde{\lambda}}{2\left( 
\tilde{\lambda}\zeta _{1}+2\alpha m_{\circ }v_{0}^{4}\right) }}.
\end{equation}%
Moreover, the probability current density 
\begin{eqnarray}
j_{x} &=&\sqrt{v_{F}\left( x\right) }\psi _{1}^{\ast }\left( x\right) \sqrt{%
v_{F}\left( x\right) }\psi _{2}\left( x\right) +c.c.  \notag \\
&=&N^{2}\tilde{\lambda}\left[ \left( i-i\right) \sin \left( \frac{\tilde{%
\lambda}}{\alpha v_{0}}\tanh \alpha x\right) \cos \left( \frac{\tilde{\lambda%
}}{\alpha v_{0}}\tanh \alpha x\right) \right]  \notag \\
&=&0,
\end{eqnarray}%
which is expected to vanish for bound states. As a result, the Dirac spinor
in (6) and the related two components $\psi _{1}\left( x\right) $ and $\psi
_{2}\left( x\right) $ represent BIC-like solution.

However, to make the unphysical solution in (19) satisfy the related
boundary conditions $\phi _{1}\left( q\right) =0$ at $q\left( x\right) =\pm
1/\alpha v_{0}$ (hence becomes a physically admissible solution) one may
shift $q\left( x\right) \longrightarrow q\left( x\right) +1/\alpha v_{0}$
and recast the solution as%
\begin{equation}
\phi _{1}\left( q\right) =\sin \left[ \tilde{\lambda}\left( q+\frac{1}{%
\alpha v_{0}}\right) \right] .
\end{equation}%
This would immediately vanish at $q=-1/\alpha v_{0}$, and yield%
\begin{equation}
\tilde{\lambda}_{n}=\frac{n\pi \alpha v_{0}}{2}\Longrightarrow E_{n}=\pm 
\sqrt{\left( \frac{n\pi \alpha v_{0}}{2}\right) ^{2}+m_{\circ }^{2}v_{0}^{4}}%
\text{ ; }n=1,2,3,\cdots ,
\end{equation}%
for $q=+1/\alpha v_{0}$. In this case, one obtains the un-normalized
components as%
\begin{equation}
\psi _{1}\left( x\right) =\sqrt{\frac{\zeta _{2}}{v_{0}}}\text{sech}\left(
\alpha x\right) \sin \left[ \frac{\tilde{\lambda}_{n}}{\alpha v_{0}}\left(
\tanh \alpha x+1\right) \right] ,
\end{equation}%
\begin{equation}
\psi _{2}\left( x\right) =-i\sqrt{\frac{\zeta _{1}}{v_{0}}}\text{sech}\left(
\alpha x\right) \cos \left[ \frac{\tilde{\lambda}_{n}}{\alpha v_{0}}\left(
\tanh \alpha x+1\right) \right] ,
\end{equation}%
and the Dirac spinor would consequently read%
\begin{equation}
\psi _{n}\left( x\right) =N_{n}\,\text{sech}\left( \alpha x\right) \left( 
\begin{tabular}{l}
$\ \ \sqrt{\frac{\zeta _{2}}{v_{0}}}\ \sin \left[ \frac{n\pi }{2}\left(
\tanh \alpha x+1\right) \right] \medskip $ \\ 
$-i\sqrt{\frac{\zeta _{1}}{v_{0}}}$ $\cos \left[ \frac{n\pi }{2}\left( \tanh
\alpha x+1\right) \right] $%
\end{tabular}%
\right) ,
\end{equation}%
Where $N_{n}$ is given in (23). Yet, it should be noticed here that for $%
m_{\circ }=0$ one may recover the final results of Panella-Roy's\ model \cite%
{13} to obtain $E_{n}=\pm n\pi \alpha v_{0}/2$, $N_{n}=\sqrt{\alpha
v_{0}/2E_{n}}$, and%
\begin{equation}
\psi _{n}\left( x\right) =\sqrt{\frac{\alpha }{2}}\,\text{sech}\left( \alpha
x\right) \left( 
\begin{tabular}{l}
$\ \ \sin \left[ \frac{n\pi }{2}\left( \tanh \alpha x+1\right) \right]
\medskip $ \\ 
$-i$ $\cos \left[ \frac{n\pi }{2}\left( \tanh \alpha x+1\right) \right] $%
\end{tabular}%
\right)
\end{equation}

As such, what may look like as a BIC-solution (documented in (19) and (24))
may turn out to be a bound state solution with discrete energy levels, if
the proper physical boundary conditions are invested in the process
(documented in (25) and (26)). Moreover, one observes a shift-up of order $%
m_{\circ }^{2}v_{0}^{4}$ in the total energy squared, $E_{n}^{2}$, and some
scaling factors in the components of the Dirac spinor (i.e., $\sqrt{\zeta
_{2}/v_{0}}$ for $\psi _{1}\left( x\right) $ and $\sqrt{\zeta _{1}/v_{0}}$
for $\psi _{2}\left( x\right) $), as discrepancies between our current model
and Panella-Roy's\ model \cite{13}. Obviously, should our $m\left( x\right)
=m_{\circ }$ (i.e., the rest mass) and $v_{F}\left( x\right) =c$ (i.e.,
speed of light), then our $q\left( x\right) =x/c$ and equation (11) would
collapse into the regular textbook Dirac equation for a free particle where
the total energy reads $E=\pm m_{\circ }c^{2}$.

\subsection{Parallel to Panella-Roy's model: $v_{F}\left( x\right)
=v_{0}\left( 1+\protect\alpha ^{2}x^{2}\right) $}

We now consider that the PD-mass as%
\begin{equation*}
m\left( x\right) =\frac{m_{\circ }}{(1+\alpha ^{2}x^{2})^{2}}
\end{equation*}%
and the PD-Fermi velocity as%
\begin{equation}
v_{F}\left( x\right) =v_{0}\left( 1+\alpha ^{2}x^{2}\right) \Longrightarrow 
\text{ }q\left( x\right) =\frac{1}{\alpha v_{0}}\arctan \alpha x.
\end{equation}%
It is easy to observe similar scenario as that associated with $\phi
_{1}\left( q\right) $ of (19), where in the current case the particle
described in (16) is now confined to move within $-\pi /2\alpha v_{0}\leq
q\left( x\right) \leq \pi /2\alpha v_{0}$. The unphysical solution then reads%
\begin{equation}
\phi _{1}\left( q\right) =\sin \tilde{\lambda}q=\sin \left( \frac{\tilde{%
\lambda}}{\alpha v_{0}}\arctan \alpha x\right) .
\end{equation}%
This would, in turn, imply that%
\begin{equation}
\psi _{1}\left( x\right) =\frac{\tilde{\psi}_{1}\left( x\right) }{\sqrt{%
v_{F}\left( x\right) }}=\ \sqrt{\frac{\zeta _{2}}{v_{0}}}\frac{1}{\sqrt{%
1+\alpha ^{2}x^{2}}}\sin \left[ \frac{\tilde{\lambda}}{\alpha v_{0}}\arctan
\alpha x\right] ,
\end{equation}%
and%
\begin{equation}
\psi _{2}\left( x\right) =\frac{\tilde{\psi}_{2}\left( x\right) }{\sqrt{%
v_{F}\left( x\right) }}=-i\ \sqrt{\frac{\zeta _{1}}{v_{0}}}\frac{1}{\sqrt{%
1+\alpha ^{2}x^{2}}}\cos \left[ \frac{\tilde{\lambda}}{\alpha v_{0}}\arctan
\alpha x\right] .
\end{equation}%
Therefore,%
\begin{equation}
\int\limits_{-\infty }^{\infty }\rho \left( x\right) dx=\int\limits_{-\infty
}^{\infty }\frac{N^{2}}{v_{0}\left( 1+\alpha ^{2}x^{2}\right) }\left[ \zeta
_{1}+2m_{\circ }v_{0}^{2}\sin ^{2}\left( \frac{\tilde{\lambda}}{\alpha v_{0}}%
\arctan \alpha x\right) \right] \,dx=1\Longrightarrow N=\sqrt{\frac{\alpha
v_{0}}{\pi \zeta _{1}+\pi m_{\circ }v_{0}^{2}}},
\end{equation}%
and%
\begin{equation}
j_{x}=N^{2}\tilde{\lambda}\left[ \left( i-i\right) \sin \left( \frac{\tilde{%
\lambda}}{\alpha v_{0}}\arctan \alpha x\right) \cos \left( \frac{\tilde{%
\lambda}}{\alpha v_{0}}\arctan \alpha x\right) \right] =0,
\end{equation}%
indicating the existence of bound states. As such, the Dirac spinor in (6)
and the related components $\psi _{1}\left( x\right) $ and $\psi _{2}\left(
x\right) $ represent a BIC-like solution.

However, the physically admissible solution would be achieved through a
shift in $q\left( x\right) \longrightarrow q\left( x\right) +\pi /2\alpha
v_{0}$ to read%
\begin{equation}
\phi _{1}\left( q\right) =\sin \left[ \tilde{\lambda}\left( q+\frac{\pi }{%
2\alpha v_{0}}\right) \right] ,
\end{equation}%
and yields%
\begin{equation}
\tilde{\lambda}_{n}=n\alpha v_{0}\Longrightarrow E_{n}=\pm \sqrt{\left(
n\alpha v_{0}\right) ^{2}+m_{\circ }^{2}v_{0}^{4}}\text{ ; }n=1,2,3,\cdots .
\end{equation}%
Therefore,%
\begin{equation}
\psi _{1}\left( x\right) =\frac{\tilde{\psi}_{1}\left( x\right) }{\sqrt{%
v_{F}\left( x\right) }}=\sqrt{\frac{\zeta _{2}}{v_{0}}}\frac{1}{\sqrt{%
1+\alpha ^{2}x^{2}}}\sin \left[ n\left( \arctan \alpha x+\frac{\pi }{2}%
\right) \right] ,
\end{equation}%
and%
\begin{equation}
\psi _{2}\left( x\right) =\frac{\tilde{\psi}_{2}\left( x\right) }{\sqrt{%
v_{F}\left( x\right) }}=-i\ \sqrt{\frac{\zeta _{1}}{v_{0}}}\frac{1}{\sqrt{%
1+\alpha ^{2}x^{2}}}\cos \left[ n\left( \arctan \alpha x+\frac{\pi }{2}%
\right) \right] .
\end{equation}%
Consequently, the Dirac spinor would read%
\begin{eqnarray}
\psi _{n}\left( x\right) &=&N_{n}\frac{1}{\sqrt{1+\alpha ^{2}x^{2}}}\left( 
\begin{tabular}{l}
$\ \ \ \sqrt{\frac{\zeta _{2}}{v_{0}}}\sin \left[ n\left( \arctan \alpha x+%
\frac{\pi }{2}\right) \right] \medskip $ \\ 
$-i$ $\sqrt{\frac{\zeta _{1}}{v_{0}}}\cos \left[ n\left( \arctan \alpha x+%
\frac{\pi }{2}\right) \right] $%
\end{tabular}%
\right) \text{; }n=1,2,3,\cdots  \notag \\
&=&\sqrt{\left( \frac{\alpha v_{0}}{\pi \zeta _{1}+\pi m_{\circ }v_{0}^{2}}%
\right) }\frac{1}{\sqrt{1+\alpha ^{2}x^{2}}}\left( 
\begin{tabular}{l}
$\ \ \ \sqrt{\frac{\zeta _{2}}{v_{0}}}\sin \left[ n\left( \arctan \alpha x+%
\frac{\pi }{2}\right) \right] \medskip $ \\ 
$-i$ $\sqrt{\frac{\zeta _{1}}{v_{0}}}$ $\cos \left[ n\left( \arctan \alpha x+%
\frac{\pi }{2}\right) \right] $%
\end{tabular}%
\right) .
\end{eqnarray}%
Moreover, for the case when $m\left( x\right) =0$ one may obtain $E_{n}=\pm 
\sqrt{\left( n\alpha v_{0}\right) ^{2}}$ and%
\begin{equation}
\psi _{n}\left( x\right) =\sqrt{\frac{\alpha }{\pi }}\frac{1}{\sqrt{1+\alpha
^{2}x^{2}}}\left( 
\begin{tabular}{l}
$\ \ \ \sin \left[ n\left( \arctan \alpha x+\frac{\pi }{2}\right) \right]
\medskip $ \\ 
$-i$ $\cos \left[ n\left( \arctan \alpha x+\frac{\pi }{2}\right) \right] $%
\end{tabular}%
\right) .
\end{equation}

Again one observes similar effects of the $m\left( x\right) v_{F}\left(
x\right) ^{2}=A$ setting on the total energy and on the components of the
Dirac spinor as those mentioned in the above example.

\section{(1+1)-Dirac P\"{o}schl-Teller holes for $v_{F}\left( x\right)
=v_{0} $ and a PDM $m(x)\neq 0$}

Let us now consider the case where $v_{F}\left( x\right)
=v_{0}\Longrightarrow q\left( x\right) =x/v_{0}$ and%
\begin{equation}
m\left( x\right) =\frac{m_{\circ }}{\sin \alpha x}.
\end{equation}%
Under such settings, the effective potential in (15) would read%
\begin{equation}
V_{eff}\left( q\right) =\frac{\alpha ^{2}v_{0}^{2}}{16}+\frac{\left(
m_{\circ }^{2}v_{0}^{4}-5\alpha ^{2}v_{0}^{2}/16\right) }{\sin ^{2}\left(
\alpha v_{0}q\right) }.
\end{equation}%
Which is obviously a shifted P\"{o}schl-Teller type periodical potential
(cf., e.g., \cite{15}). In this case, one may rewrite (14) as%
\begin{equation}
-\partial _{q}^{2}\phi _{1}\left( q\right) +\frac{\tilde{V}_{\circ }}{2}%
\left[ \frac{s(s-1)}{\sin ^{2}\left( \alpha v_{0}q\right) }\right] \phi
_{1}\left( q\right) =\left( E^{2}-\frac{\alpha ^{2}v_{0}^{2}}{16}\right)
\phi _{1}\left( q\right) ,
\end{equation}%
where $\tilde{V}_{\circ }=2\alpha ^{2}v_{0}^{2}$ and $s(s-1)=m_{\circ
}^{2}/\alpha ^{2}-5/16$. Such periodical potential setting imposes infinite
impenetrable barriers manifested by the singularities between the holes
(i.e., at $q=0,\pi /v_{0}\alpha ,2\pi /v_{0}\alpha ,\cdots $ or equivalently
at $x=0,\pi /\alpha ,2\pi /\alpha ,\cdots $.). Here we pick up the hole
within $0\leq x=v_{0}q\leq \pi /\alpha $ to obtain%
\begin{equation}
\phi _{1}\left( q\right) =\,\sin ^{s}\left( \alpha v_{0}q\right)
\,_{2}F_{1}\left( -n,s+n,s+\frac{1}{2};\sin ^{2}\left( \alpha v_{0}q\right)
\right)
\end{equation}%
and%
\begin{equation}
E_{n}=\pm \frac{\alpha v_{0}}{4}\sqrt{1+16(s+2n)^{2}}\text{ ; }%
n=0,1,2,\cdots .
\end{equation}%
Where%
\begin{equation}
s=\frac{1}{2}+\sqrt{m_{\circ }^{2}\alpha ^{2}-\frac{1}{16}}>1
\end{equation}%
Then, one would, in a straightforward manner, cast%
\begin{equation}
\psi _{1}\left( x\right) =N_{n}\sqrt{\frac{E_{n}}{v_{0}}+\frac{m_{\circ
}v_{0}}{\sin \alpha x}}\sin ^{s}\left( \alpha x\right) \,_{2}F_{1}\left(
-n,s+n,s+\frac{1}{2};\sin ^{2}\left( \alpha x\right) \right)
\end{equation}%
and find $\psi _{2}\left( x\right) $, using (7), to construct the Dirac
spinor of (5). Obviously, BIC-like bound states are not feasible here and
only discrete bound state solutions are obtained.

\section{(1+1)-Dirac effective harmonic oscillator toy: a by-product of $%
v_{F}\left( x\right) =v_{0}x$ and a $m(x)=A/x$}

Consider a singular position-dependent mass along with a linear
position-dependent Fermi velocity of the forms%
\begin{equation}
m(x)=A/x\text{; }v_{F}\left( x\right) =v_{0}x\text{ ; }x\in \left( 0.\infty
\right) \Longrightarrow x=e^{v_{0}q}\Longrightarrow q\left( x\right) =\ln
x^{1/v_{0}};\text{ }q\in \left( -\infty ,\infty \right) .
\end{equation}%
In this case, equation (14) along with (15) would yield%
\begin{equation}
-\partial _{q}^{2}\phi _{1}\left( q\right) +\frac{1}{4}\omega ^{2}q^{2}\phi
_{1}\left( q\right) =\left( E^{2}+\frac{v_{0}^{2}}{16}\right) \phi
_{1}\left( q\right) ,
\end{equation}%
with $\omega =2Av_{0}^{3}$. Obviously, $\phi _{1}\left( q=\pm \infty \right)
=0$ represent the boundary conditions for the current Dirac harmonic
oscillator at hand. In a straightforward manner, one would use the
traditional textbook procedure and find that%
\begin{equation}
E^{2}+\frac{v_{0}^{2}}{16}=\omega \left( n+\frac{1}{2}\right)
\Longrightarrow E_{n}=\pm \sqrt{Av_{0}^{3}\left( 2n+1\right) -\frac{v_{0}^{2}%
}{16}}.
\end{equation}%
and%
\begin{equation}
\phi _{1,n}\left( q\right) =e^{-Av_{0}^{3}q^{2}/2}H_{n}\left( \sqrt{%
Av_{0}^{3}}q\right) ,
\end{equation}%
where $H_{n}\left( \sqrt{Av_{0}^{3}}q\right) $ are the Hermite polynomials.
Then we may obtain%
\begin{equation}
\psi _{1,n}\left( x\right) =N_{n}\sqrt{\frac{E_{n}}{v_{0}x}+Av_{0}}e^{-\frac{%
Av_{0}}{2}\ln ^{2}x}H_{n}\left( \sqrt{Av_{0}}\ln x\right) ,
\end{equation}%
and find $\psi _{2}\left( x\right) $ using (7) to construct the Dirac spinor
of (5). Only discrete bound state solutions are observed here.

\section{Concluding remarks}

In this work, we have considered the (1+1)-Dirac particles where the mass
and the Fermi velocity are both position-dependent. An alternative
methodical proposal is proposed in such a way that the Panella-Roy's\ model 
\cite{13} becomes a special case. The set of $m\left( x\right) $ and $%
v_{F}\left( x\right) $ that satisfies $m\left( x\right) v_{F}\left( x\right)
^{2}=A$ is a wider set than that used by Panella and Roy who have used
massless Dirac particles. Moreover, analogous to the well known textbook
non-relativistic limit for Dirac particles (i.e., rest mass energy $m_{\circ
}c^{2}>>E_{bind}$, where $E_{bind}=E-m_{\circ }c^{2}$), we have used the
limit where $m\left( x\right) v_{F}\left( x\right) ^{2}>>E_{bind}$ for
non-zero PD-masses. To the best of our knowledge such methodical proposal
has not been reported elsewhere.

For Dirac particles with $m\left( x\right) $ and $v_{F}\left( x\right) $
satisfying $m\left( x\right) v_{F}\left( x\right) ^{2}=A$, we have reported
feasible BIC-like and discrete bound-states solutions (documented in section
III). They are in an almost exact accord with the scenario reported in the
Panella-Roy's model. However, we have also observed a shift-up of order $%
m_{\circ }^{2}v_{0}^{4}=A^{2}$ in the total energy squared, $E_{n}^{2}$, and
some scaling factors in the components of the Dirac spinor (i.e., $\sqrt{%
\zeta _{2}/v_{0}}$ for $\psi _{1}\left( x\right) $ and $\sqrt{\zeta
_{1}/v_{0}}$ for $\psi _{2}\left( x\right) $). Moreover, the results of our
methodical proposal collapse into those of Panella and Roy in \cite{13} for $%
m_{\circ }=0$. Yet, should one use $m\left( x\right) =m_{\circ }$ (i.e., the
rest mass) and $v_{F}\left( x\right) =c$ (i.e., speed of light), then $%
q\left( x\right) =x/c$ and equation (11) would collapse into the regular
textbook Dirac equation for free particle, where the total energy is $E=\pm
m_{\circ }c^{2}$.

Finally, for the case where $m\left( x\right) v_{F}\left( x\right) ^{2}\neq
A $, we have shown that Dirac particles may be trapped in an effective force
fields produced by both their PD-mass and PD-Fermi velocity. This is
documented in the effective P\"{o}schl-Teller and the effective harmonic
oscillator models discussed in sections IV and V, respectively. No BIC-like
bound state solutions are observed for these models.

\end{document}